\documentclass[pra,twocolumn,showpacs]{revtex4}
\usepackage{graphicx,graphics,psfrag,amsmath,calc,amssymb}

\begin{document}

\title{Characteristics of Bose-Einstein condensation in an optical lattice}

\author{G.-D. Lin, Wei Zhang, and L.-M. Duan}

\affiliation{FOCUS Center and MCTP, Department of Physics,
University of Michigan, Ann Arbor, Michigan 48109, USA}
\date{\today}

\begin{abstract}
We discuss several possible experimental signatures of the Bose-Einstein
condensation (BEC) transition for an ultracold Bose gas in an inhomogeneous
optical lattice. Based on the commonly used time-of-flight imaging
technique, we show that the momentum-space density profile in the first
Brillouin zone, supplemented by the visibility of interference patterns,
provides valuable information about the system. In particular, by crossing
the BEC transition temperature, the appearance of a clear bimodal structure
sets a qualitative and universal signature of this phase transition.
Furthermore, the momentum distribution can also be applied to extract the
condensate fraction, which may serve as a promising thermometer in such a
system.
\end{abstract}

\maketitle

%
%

\section{Introduction}

\label{sec:introduction}

There has been significant interest in ultracold atomic gases in optical
lattices, partly stimulated by the possibility of simulating strongly
correlated many-body systems~\cite{bloch-05}. With extraordinary
controllability, ultracold atomic gases in lattices provide a promising
experimental platform to help tackling many important problems in
multidisciplinary fields. Among these topics, the emergence of condensation
and superfluid order in an optical lattice, and how the superfluid order
transforms into other ordered states, is a problem which catches great
attention over the past decade. With current technology, condensation and
superfluidity are obtained for both Bose and Fermi gases in optical
lattices, so that various phase transitions can be investigated~\cite%
{greiner-02, ketterle}.

To experimentally investigate these phase transitions, present techniques
with ultracold atomic gases heavily rely on detection based on the
time-of-flight imaging, where the interference pattern and its visibility
are suggested to be signatures of Bose condensation within an optical
lattice~\cite{greiner-02}. Recent studies show that even for a thermal
lattice gas above the BEC transition temperature $T_{c}$, interference peaks
with observable visibility are still present~\cite{diener-07, gerbier-07,
yi-07}. This thermal visibility can be large for an ideal Bose gas in a
homogeneous lattice, which could make the condensation signal ambiguous~\cite%
{diener-07}. However, for practical systems with atomic interaction and an
inhomogeneous global trap, the thermal visibility becomes significantly
smaller~\cite{gerbier-07,yi-07}, and the appearance of sharp interference
peaks is still associated with the BEC transition. It is also suggested that
the bimodal structure of the atomic momentum distribution in the first
Brillouin zone, combined with the interference peaks, provides an additional
unambiguous signal for the Bose condensation in an optical lattice~\cite%
{yi-07}. A very recent experiment has used the onset of the
bi-modal distribution and the associated condensate fraction to
identify the superfluid-to-Mott-insulator transition point \cite%
{NIST}.

In this manuscript, we provide a detailed study of a Bose gas in a
three-dimensional (3D) inhomogeneous optical lattice, both below and above
the BEC transition temperature. We discuss several properties including the
visibility, the width of the interference peak, and the momentum
distribution of the resulting interference pattern. The main results are as
follows. First, all the quantities mentioned above can characterize the BEC
transition for the experimental systems with interacting atoms in an
inhomogeneous optical lattice. The large thermal visibility applies only to
some particular parameters which are not directly responsible for current
experiments. In the case when the thermal visibility is large, a substantial
variation of the peak width or the appearance of a bimodal structure for the
atomic momentum distribution may work as a better signature for the
condensation transition. Second, below the BEC transition temperature, the
visibility and the peak width become insensitive to the system temperature,
hence can not be applied as a practical thermometer. To fulfill this gap,
the bimodal structure of the atomic momentum distribution gives a way to
extract the condensate fraction through the bimodal fitting. The resulting
condensate fraction provides a sensitive indicator of the system
temperature, hence may serve as a potential thermometer for this important
system.

The calculation techniques in this paper are similar to what we present in
Ref.~\cite{yi-07}. The remainder of this manuscript is organized as follows.
In Sec.~\ref{sec:ideal}, we first consider the situation of free bosons in
an optical lattice within a global harmonic trap, and investigate the
general behavior of the interference visibility and the atomic momentum
distribution. In the absence of interaction, the problem is significantly
simplified such that exact solutions are available. These exact solutions,
on the one hand, are valuable for qualitative understanding of the system
and its properties, and on the other hand, could be directly compared with
experiments when the Feshbach resonance technique is applied to turn off the
atomic interaction. After studying the free Bose gas, we then extend our
discussion in Sec.~\ref{sec:inter} to the case of interacting bosons, where
effects of the global harmonic trap and the interaction have to be taken
into account together. In order to deal with the trap, we adapt the local
density approximation (LDA), which works well when the interaction energy
scale is significantly larger than the trapping energy scale (this condition
is typically valid for current experiments). Restricting our discussion to
weakly interacting bosons that are away from the Mott region, we apply the
Hartree-Fock-Bogoliubov-Popov (HFBP) approximation to deal with the atomic
interaction~\cite{andersen-04, vanOosten-01, rey-03}. The HFBP method can
provide a reliable description except for a narrow region around the BEC
transition temperature~\cite{griffin-98,andersen-04}. Taking into account
the effect of the global trap, this questionable region only corresponds to
a thin shell in three dimensions and its influence to the global properties
is small. Therefore, we expect that the HFBP method can give reliable
results for the atomic momentum distribution and the condensate fraction.


\section{Ideal Bose gas in an inhomogeneous optical lattice}

\label{sec:ideal}

In this section, we discuss an ideal Bose gas in an inhomogeneous optical
lattice with a global harmonic trap. For completeness, we briefly review the
formalism~\cite{yi-07} before presenting various calculation results. We
consider the atoms in a cubic lattice with an additional spherically
symmetric harmonic trap~\cite{note1}. The Hamiltonian takes the form %
\begin{equation}
H=\int d^{3}\mathbf{r}\Psi ^{\dagger }(\mathbf{r})\left[ -\frac{\hbar
^{2}\nabla _{\mathbf{r}}^{2}}{2m}+V_{\mathrm{op}}(\mathbf{r})+V(\mathbf{r})%
\right] \Psi (\mathbf{r}),  \label{eqn:Hamiltonian-non}
\end{equation}
where $\Psi $ represents the bosonic field operator, $m$ is the atomic mass,
$V_{\mathrm{op}}(\mathbf{r})\equiv V_{0}\sum_{i=x,y,z}\sin ^{2}(\pi r_{i}/d)$
is the optical lattice potential with lattice spacing $d$, and $V(\mathbf{r}%
)\equiv m\omega ^{2}r^{2}/2$ is the global harmonic trapping potential. In
practice, the global harmonic trap $V(\mathbf{r})$ typically varies much
slower than the optical lattice potential $V_{\mathrm{op}}(\mathbf{r})$, so
the Hamiltonian can be separated into two parts with fast and slow
variations, respectively. The fast-varying part can be diagonalized by
introducing the expansion of bosonic field operators %
\begin{equation}
\Psi (\mathbf{r})=\sum_{\mathbf{R}}w(\mathbf{r}-\mathbf{R})a_{\mathbf{R}},
\label{eqn:expansion}
\end{equation}
where $w(\mathbf{r})$ is the Wannier function associated with the lattice
potential $V_{\mathrm{op}}(\mathbf{r)}$, $a_{\mathbf{R}}$ is the
annihilation operator on site $\mathbf{R}$, and the summation is over all
lattice sites. After transforming to the momentum space, the Fourier
components of $\Psi (\mathbf{r})$, $w(\mathbf{r})$, and $a_{\mathbf{R}}$
satisfy the following relation %
\begin{equation}
\Psi (\mathbf{k})=w(\mathbf{k})a_{\mathbf{k}}.  \label{eqn:expansion2}
\end{equation}

Representing the fast and slow varying components of $H$ in terms of $a_{%
\mathbf{k}}$ and $a_{\mathbf{R}}$, respectively, the original Hamiltonian
Eq.~(\ref{eqn:Hamiltonian-non}) can be written as %
\begin{equation}
H=\sum_{\mathbf{k}\in \mathrm{1BZ}}\epsilon _{\mathbf{k}}a_{\mathbf{k}%
}^{\dagger }a_{\mathbf{k}}+\sum_{\mathbf{R}}V(\mathbf{R})a_{\mathbf{R}%
}^{\dagger }a_{\mathbf{R}},  \label{eqn:hamiltonian-non2}
\end{equation}
where the summation over quasi-momentum $\mathbf{k}$ is restricted to the
first Brillouin zone (1BZ). Here, we assume that the lattice depth $V_{0}$
is strong enough such that the band gap is large and atoms are confined to
the lowest band with dispersion relation $\epsilon _{\mathbf{k}%
}=-2t\sum_{i=x,y,z}\cos (k_{i}d)$. The tunneling rate $t$ can be well
estimated by $t\approx (3.5/\sqrt{\pi })V_{0}^{3/4}\exp (-2\sqrt{V_{0}})$,
where the recoil energy $E_{\mathrm{R}}\equiv \hbar ^{2}\pi ^{2}/(2md^{2})$
is used as the energy unit~\cite{duan-05}.

In principle, the resulting Hamiltonian Eq. (\ref{eqn:hamiltonian-non2}) can
be directly diagonalized for arbitrary $V(\mathbf{R})$. However, the
numerical calculation is usually very heavy in three dimensions due to the
presence of a large number of lattice sites. In this case, the
diagonalization process can be significantly simplified by noticing that the
indices $\mathbf{k}$ and $\mathbf{R}$ in Eq. (\ref{eqn:hamiltonian-non2})
are reminiscent of the coordinate and the momentum variables in quantum
mechanics. This observation allows us to write down a first quantization
Hamiltonian corresponding to Eq. (\ref{eqn:hamiltonian-non2}) in the
momentum space, where $\mathbf{R}$ is replaced by the momentum gradient $%
-i\nabla _{\mathbf{k}}$~\cite{polkovnikov-02,hooley-04,rey-05,yi-07}. The
resulting effective Hamiltonian thus takes the form %
\begin{equation}
H_{\mathrm{eff}}=-\frac{1}{2}m\omega ^{2}\nabla _{\mathbf{k}}^{2} +\epsilon_{%
\mathbf{k}},  \label{eqn:hamiltonian-non3}
\end{equation}
which represents free bosons with effective mass $m^{\ast }\equiv \hbar
^{2}/(m\omega ^{2})$ in a periodic potential $\epsilon _{\mathbf{k}}$ with
period $|\mathbf{G}|=2\pi /d$ along all three principal directions.
Furthermore, since $\nabla _{\mathbf{k}}^{2}$ and $\epsilon _{\mathbf{k}}$
are separable, this Hamiltonian can be reduced to three one-dimensional
problems which require much less effort to solve. Notice that the properties
of this effective Hamiltonian depend only on the ratio of $\hbar^{2}\omega
^{2}/(t E_{\mathrm{R}})$, which suggests that the variation of $\omega$ and $%
V_{0}$ can be scaled to each other by fixing the dimensionless parameter $%
\hbar ^{2}\omega ^{2}/(V_{0}^{3/4} e^{-2\sqrt{V_{0}}})$, where $E_{\mathrm{R}%
}$ is used as the energy unit. In this section, we will keep $V_{0}$ fixed ($%
V_{0}=10E_{\mathrm{R}}$) and look at changes of the system properties under
variation of the global trapping frequency $\omega$. With different lattice
barriers $V_{0}$, one can directly read out the result by simply re-scaling $%
\omega $ to keep $\hbar ^{2}\omega ^{2}/(V_{0}^{3/4} e^{-2\sqrt{V_{0}}})$
fixed (a smaller barrier thus corresponds to a larger effective trapping
frequency).

The quasi-momentum distribution $\langle a_{\mathbf{k}}^{\dagger }a_{\mathbf{%
k}}\rangle $ is then given by the square of the eigenstate wave functions $%
\phi _{\mathbf{n}}(\mathbf{k})$ of $H_{\mathrm{eff}}$, where the expectation
value is obtained by averaging over all eigenstates $\mathbf{n}$ with a Bose
distribution factor $g(E_{\mathbf{n}})=1/\exp [\beta (E_{\mathbf{n}}-\mu
)-1] $. Here, $E_{\mathbf{n}}$ is the corresponding eigenenergy, $\mu $ is
the chemical potential, and $\beta =1/(k_{B}T)$ is the inverse temperature.
Taking into account the presence of the Wannier function in Eq. (3), the
atomic \textit{real} momentum distribution is %
\begin{eqnarray}
n(\mathbf{k}) &=&\langle \Psi ^{\dagger }(\mathbf{k})\Psi (\mathbf{k}%
)\rangle =|w(\mathbf{k})|^{2}\langle a_{\mathbf{k}}^{\dagger }a_{\mathbf{k}%
}\rangle  \notag  \label{eqn:mom-dis-non} \\
&=&|w(\mathbf{k})|^{2}\sum_{\mathbf{n}}g(E_{\mathbf{n}})|\phi _{\mathbf{n}}(%
\mathbf{k})|^{2}.
\end{eqnarray}
For a free gas, this momentum distribution remains unchanged during
expansion, so that the signal from the time-of-flight image taking along a
crystallographic axis, say $z$ direction, is just the columnar density %
\begin{equation}
n_{\perp }(k_{x},k_{y})=\int n(\mathbf{k})dk_{z}.  \label{eqn:col-den}
\end{equation}
\begin{figure}[tbp]
\begin{center}
\includegraphics[width=8.0cm]{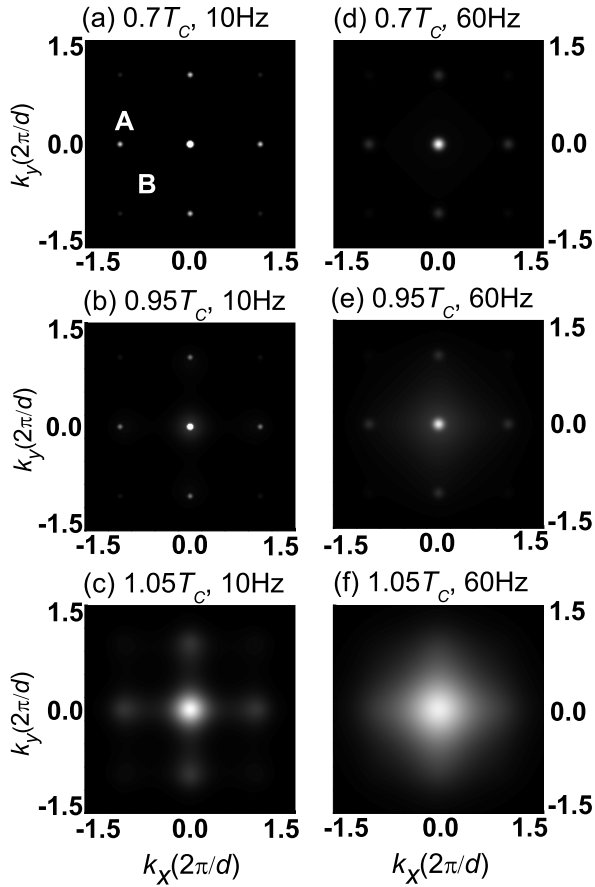}
\end{center}
\caption{Calculated column-integrated momentum density profile near the
first Brillouin zone measured through the time-of-flight imaging, taking for
two different trapping frequencies at various temperatures. Parameters
chosen in these plots are close to those for a typical experiment of $^{87}$%
Rb, with lattice depth $V_{0}=10E_{\mathrm{R}}$ and a total number of
particles $N=10^{5}$. The global harmonic trapping frequencies are $\protect%
\omega =2\protect\pi \times 10$ Hz for (a)-(c), and $2\protect\pi \times 60$
Hz for (d)-(f). }
\label{fig:image-non}
\end{figure}

Using the technique sketched above, we show in Fig.~\ref{fig:image-non} the
calculation results for the column-integrated momentum distribution as
record by the time-of-flight images, both below and above the BEC transition
temperature $T_{c}$. In these plots, we choose parameters close to those for
a typical experiment of $^{87}$Rb atoms, where the lattice depth $%
V_{0}=10E_{\mathrm{R}}$, and the total number of particles $N = 10^{5}$. For
this finite system, the transition temperature $T_{c}$ is determined by
requiring that the number of atoms in the ground state is of the order of $1$
when $T>T_{c}$ and increases by orders of magnitude when $T$ crosses $T_{c}$%
. Since the total atom number satisfies $N\gg 1$, the transition is actually
very sharp, and $T_{c}$ is well defined by the requirement above.

>From Fig.~\ref{fig:image-non}, one can see that for the case of a very weak
global trap ($\omega =2\pi \times 10$ Hz), the interference peaks are indeed
clearly visible even above the transition temperature. However, the BEC
transition is still evident from the time-of-flight images as the
interference peaks become much sharper when $T$ gets below $T_{c}$. For the
case with a stronger global trap ($\omega =2\pi \times 60$ Hz, which is
close to the value in experiments), the interference peaks become blurred
when $T>T_{c}$ (although one can still read some pattern). Again, across the
transition temperature $T_{c}$, the time-of-flight image, in particular the
sharpness of the central peak, undergoes a dramatic change.

These figures show that qualitatively a Bose condensation transition should
be visible with the time-of-flight images. To have a more quantitative
description, however, it is desirable to have some single-value indicators
which change sharply across the BEC transition so that one can characterize
this phase transition by measuring the indicators. As possible candidates,
we next discuss in detail two quantities, including the visibility of the
interference pattern and the peak width associated with the atomic momentum
distribution in Sec.~\ref{sec:vis-non} and~\ref{sec:distribution-non},
respectively. While both of the two quantities can signify the BEC
transition in a reasonably strong global trap, the peak width becomes more
accurate when the global trap gets weaker. After the condensation
transition, both of these two indicators become very insensitive to the
variation of the system temperature, so they do not provide a good
thermometer. Instead, we suggest to measure the condensate fraction from the
bimodal fitting to the central interference peak, as discussed in Sec.~\ref%
{sec:frac-non}. The measured condensate fraction change continuously with
the temperature, thus gives a good indicator for estimation of temperature
in this important system.

\subsection{Visibility of the interference pattern}

\label{sec:vis-non}

The visibility of the interference pattern has been introduced in Ref.~\cite%
{gerbier-05}. It is defined as the intensity contrast of two characteristic
points on the interference pattern~\cite{gerbier-05} %
\begin{equation}
v=\frac{n_{\perp }^{A}-n_{\perp }^{B}}{n_{\perp }^{A}+n_{\perp }^{B}},
\label{eqn:visibility}
\end{equation}%
where $n_{\perp }^{A}$ and $n_{\perp }^{B}$ are (column-integrated) atomic
intensities at sites $A$ and $B$, respectively, as shown in Fig.~\ref%
{fig:image-non}(a). The point $A$ represents the position of the secondary
peak while $B$ is along the circle of the secondary peaks where the
intensity takes its minimum. The visibility defined in this way is clearly
independent of the Wannier function [the pre-factor in Eq. (3)].

\begin{figure}[tbp]
\begin{center}
\includegraphics[width=8.0cm]{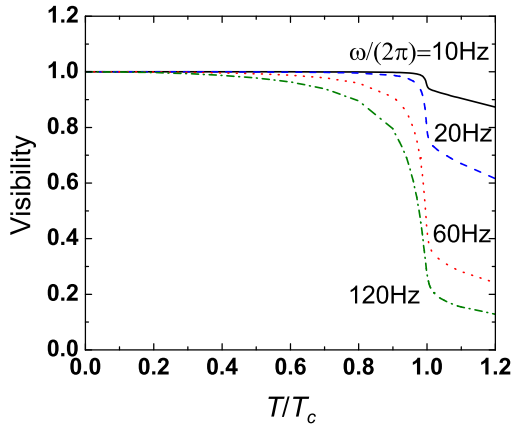}
\end{center}
\caption{(color online) The visibility as a function of temperature around the BEC
transition temperature $T_c$ for various values of trapping frequencies.
Other parameters are the same as those used in Fig.~\protect\ref%
{fig:image-non} with $V_0 = 10 E_{\mathrm{R}}$ and $N = 10^{5}$. Remind that
the variation of $V_0$ is equivalent to that of $\protect\omega$, with a
fixed value of $\hbar^2 \protect\omega^2/(V_0^{3/4} e^{-2\protect\sqrt{V_0}})
$. }
\label{fig:visi-non}
\end{figure}

The temperature dependence of the visibility $v$, especially by crossing the
BEC transition, is shown in Fig.~\ref{fig:visi-non} for various values of
trapping frequencies. For a very weak trap ($\omega =2\pi \times 10$ Hz),
the visibility is pretty high ($v>0.8$) even with the temperature
considerably larger than $T_{c}$. So in the limit of a vanishing $\omega $,
this is consistent with the results in Ref. \cite{diener-07} for free bosons
in a homogeneous lattice (without the global trap). However, for large
trapping frequencies, as pointed out in Ref.~\cite{gerbier-07}, the
visibility becomes significantly smaller when $T>T_{c}$, leading to a more
substantial drop across the BEC transition. With a trapping frequency around
$\omega =2\pi \times 60$ Hz, the visibility jumps should be pretty evident
to observe, as shown in Fig.~\ref{fig:visi-non}. However, the transition is
stretched over a wide range of temperatures (the visibility begins to drop
starting from a temperature significantly below $T_{c}$, see Fig.~\ref%
{fig:visi-non}), which may make the determination of the transition point
from the visibility less accurate. Notice that with the scaling relation, a
smaller barrier $V_0$ corresponds to a larger effective trapping frequency $%
\omega$. So with a shallower lattice, the change in the visibility across
the BEC transition gets larger for free bosons.

\begin{figure}[tbp]
\begin{center}
\includegraphics[width=7.5cm]{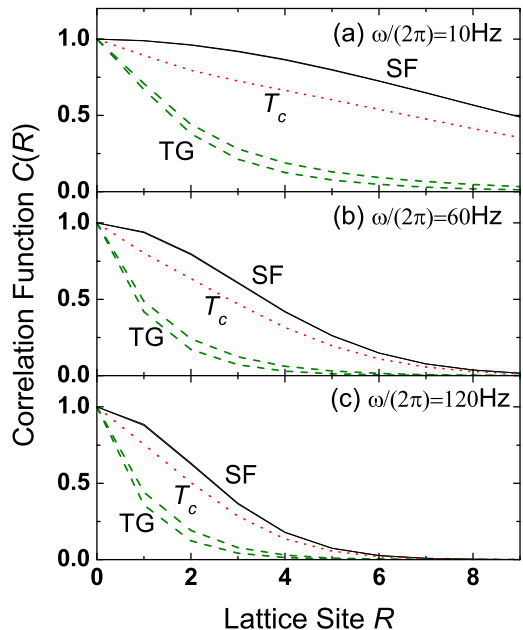}
\end{center}
\caption{(color online) Correlation functions around the trap center indicating real space
coherence below (for a superfluid, SF) and above (for a thermal gas, TG) the
BEC transition temperature, with trapping frequencies (a) $\protect\omega =2%
\protect\pi \times 10$ Hz, (b) $2\protect\pi \times 60$ Hz, and (c) $2%
\protect\pi \times 120$ Hz. Curves in each panel are taken at, from top to
bottom, $0.9T_{c}$(solid), $0.95T_{c}$(solid), $1.0T_{c}$(dotted), $1.05T_{c}
$(dashed), and $1.1T_{c}$(dashed), respectively. Parameters used in these
plots are $V_{0}=10E_{\mathrm{R}}$ and $N=10^{5}$.
For reference, the characteristic length of the single particle
ground state wavefunction adjusted by effective mass ($L \equiv \sqrt{\hbar/m^* \omega}$)
are (a) $L=23.1$, (b) $L=9.4$, and (c) $L=6.7$, respectively.}
\label{fig:corr-non}
\end{figure}

For the system temperature above $T_{c}$, there is no long range coherence
in the atomic cloud, so the finite visibility of the interference pattern is
induced by residue short-range thermal correlations. To understand the
different behavior of the visibility, we calculate the short-range thermal
correlation function around the trap center with different global trapping
potentials. The real space correlation function is defined as %
\begin{equation}
C(\mathbf{R})=\frac{\langle \Psi ^{\dagger }(\mathbf{0}) \Psi(\mathbf{R}%
)\rangle } {\sqrt{\langle \Psi ^{\dagger }(\mathbf{0}) \Psi (\mathbf{0}%
)\rangle \langle \Psi ^{\dagger }(\mathbf{R}) \Psi (\mathbf{R})\rangle }},
\label{eqn:corr-non}
\end{equation}
with $\mathbf{R}=\mathbf{0}$ indicating the trap center. In Fig.~\ref%
{fig:corr-non}, we show the correlation functions both below and above $%
T_{c} $ for different trapping frequencies. Notice that for weak trapping
potentials [see, e.g., Fig.~\ref{fig:corr-non}(a)], the correlation function
extends to several lattice sites at temperatures above $T_{c}$, indicating
the presence of thermal short-range coherence. As the trapping frequency
increases, the correlation length for a thermal gas decreases as presented
in Fig.~\ref{fig:corr-non}(b) and \ref{fig:corr-non}(c), which is consistent
with the disappearance of interference peaks as shown in Fig.~\ref%
{fig:visi-non}(f). %

\subsection{Momentum-space density profile and the peak width}

\label{sec:distribution-non}

Up to now, we discuss the visibility characterizing the contrast of the
interference pattern. In this subsection, we introduce another single-value
quantity, the peak width, which characterizes the sharpness of the
interference peak. We notice that while the visibility does not undergoes a
sudden change across the BEC transition when the global trap is weak, the
width of the central interference peak, however, always shrinks sharply when
the condensation takes place. Therefore, the peak width is always a good
indicator of the BEC transition independent of the strength of the global
trap

\begin{figure}[tbp]
\begin{center}
\includegraphics[width=8.0cm]{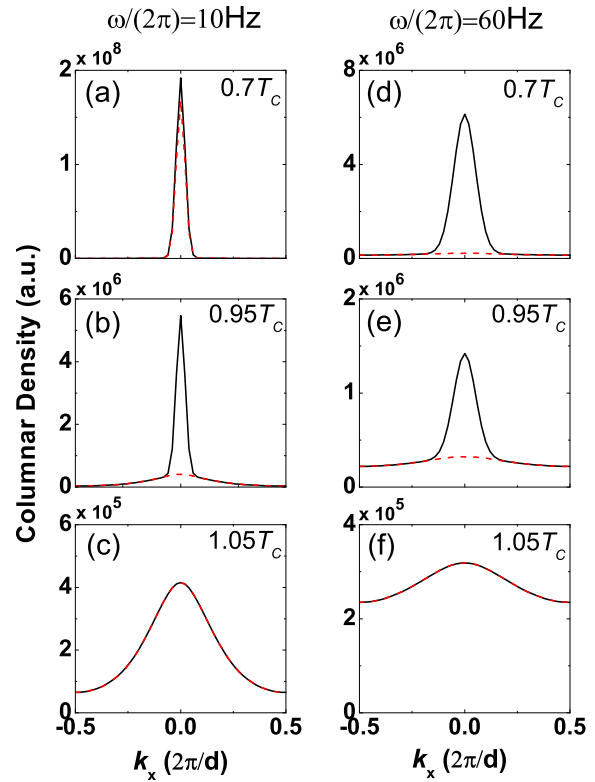}
\end{center}
\caption{(color online) Momentum space columnar density along the $x$-axis in the first
Brillouin zone. The left and right panels correspond to the cases of
trapping frequencies $\protect\omega =2\protect\pi \times 10$Hz and $2%
\protect\pi \times 60$Hz, respectively. Notice that while a thermal
distribution is present above $T_{c}$ [(c) and (f)], a clear bimodal
structure starts to appear for temperatures slightly below $T_{c}$ with only
precentral condensate fractions [$n_{0}\sim 0.14$ in (b) and $\sim 0.09$ in
(e)]. Here, the solid curves represent the total momentum density profile,
and the dashed curves represent the momentum density profile of the normal
component. Parameters used in these plots are $V_{0}=10E_{\mathrm{R}}$ and $%
N=10^{5}$. }
\label{fig:distri-non}
\end{figure}

To introduce the peak width, first we look at the atomic momentum
distribution, which gives more detailed information about the system. From
Eq. (3), the momentum distributions in other Brillouin zones are simply
copies of the distribution in the first Brillouin zone weighted by the given
Wannier function, so it suffices to study the atomic momentum profile in the
first Brillouin zone. In Fig.~\ref{fig:distri-non}, we plot the
column-integrated momentum distribution along one crystallographic axis
(e.g., the $x$-axis) passing through the center of the first Brillouin zone.
It is clear that while the density profile is a thermal distribution when $%
T>T_{c}$, a bimodal structure starts to appear when a non-zero condensate
fraction emerges in the lattice, characterized by a sharp peak at the center
of the momentum space surrounded by a flat thermal distribution. The signal
of this structural change is significant as soon as the system crosses the
BEC transition [see Fig.~\ref{fig:distri-non}(b) and (e)]. %

\begin{figure}[tbp]
\begin{center}
\includegraphics[width=8.0cm]{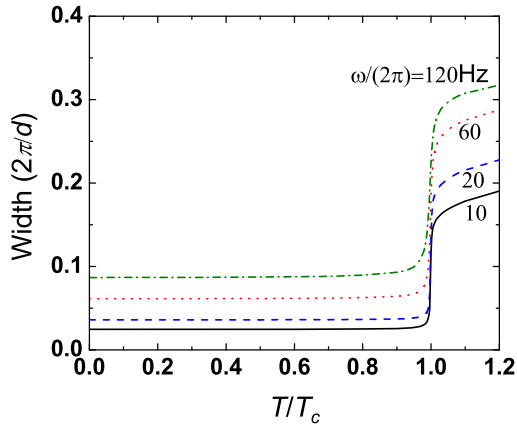}
\end{center}
\caption{(color online) The peak width (defined in the text) within the first Brillouin
zone, taking across the transition temperature for various values of
trapping frequencies. Notice that the sharp and substantial change around $%
T_{c}$ occurs for all cases hence serves as a distinctive signature of BEC
transition. Parameters used in these plots are $V_{0}=10E_{\mathrm{R}}$ and $%
N=10^{5}$. }
\label{fig:width-non}
\end{figure}

In order to characterize the sharpness of the atomic momentum distribution
in the first Brillouin zone, we introduce the peak width as a single-value
parameter. For this purpose, we first define the middle value of $%
n_{\perp}(k_{x},k_{y})$ within the first Brillouin zone %
\begin{equation}
n_{\mathrm{mid}}\equiv \frac{1}{2}\left[ \max_{(k_{x},k_{y})\in \mathrm{1BZ}%
} +\min_{(k_{x},k_{y})\in \mathrm{1BZ}}\right] n_{\perp }(k_{x},k_{y}).
\label{eqn:mid-height}
\end{equation}
In the simplest term, the peak width $w$ is measured as the radius in the
momentum space where the (column-integrated) atomic density $%
n_{\perp}(k_{x},k_{y})$ first falls to this middle value $n_{\mathrm{mid}}$.
In Fig.~\ref{fig:width-non}, we show the peak width as a function of
temperatures for various trapping frequencies. Notice that the central peak
width decreases monotonically with temperature, and most importantly,
undergoes a sharp and substantial change by crossing the transition
temperature. This distinctive feature is universal for all trapping
frequencies, hence can provide a clear criterion for the phase transition.
However, as will be discussed later, the sharpness of the change of central
peak width around $T_{c}$ is guaranteed only for ideal Bose gases. In the
presence of atomic interaction, the variation of central peak width may be
more flat and extended over a wider range of temperatures. %

\subsection{The condensate fraction as a measure of temperature}

\label{sec:frac-non}

\begin{figure}[tbp]
\begin{center}
\includegraphics[width=8.0cm]{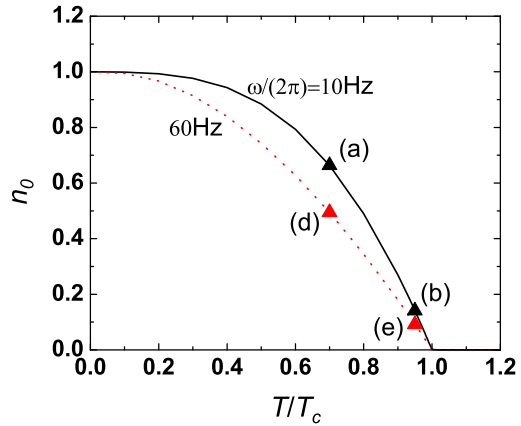}
\end{center}
\caption{(color online) The condensate fraction as a function of temperature below the BEC
transition. By extracting $n_{0}$ from the momentum distribution as shown in
Fig.~\protect\ref{fig:distri-non}(a-b) and (d-e), temperatures of such cases
can be determined correspondingly (triangles). Parameters used in this
figure are $V_{0}=10E_{\mathrm{R}}$ and $N=10^{5}$. }
\label{fig:n0-non}
\end{figure}

>From the discussion above, we notice that after the condensation
transition, both the visibility and the peak width become almost flat to
variation of temperature, as one can see from Figs. \ref{fig:visi-non} and
\ref{fig:width-non}. This means that it is hard to get any information
about the temperature of the system from the measured values of
visibility and peak width. Since temperature is one of the most
important quantity for the thermodynamical property of the system, it is
desirable to have some experimentally measurable indicator which gives a
good estimate of the temperature. The momentum density profile in principle
gives a lot of information, but it is not a single-value quantity which
makes it hard to support a direct comparison at different temperatures or
for different systems. To overcome this drawback, we note that below $T_c$,
the atomic momentum distribution in the first Brillouin zone always shows a
bimodal structure, which actually gives a universal signal for the
condensation transition. Furthermore, from the measured momentum density
profile, one can always do a bimodal fitting to figure out the atomic
fractions in the condensate and in the thermal parts, respectively. The
measured condensate fraction thus can serve as a good estimate for the
system temperature. In Fig.~\ref{fig:n0-non}, we show the calculated
condensate fraction as a function of temperature for this system with two
different global trapping frequencies. Notice that the condensate fraction $%
n_{0}$ changes steadily and monotonically as the temperature $T$ varies.
>From the relation $n_{0}(T)$, one can estimate the temperature $T$ through
the experimentally measurable $n_{0}$. Therefore, the condensate fraction $%
n_{0}$ gives a single-value quantity which can serve as a criterion for the
condensation transition (with $n_{0}>0$), as well as an indicator of the
system temperature. %

\section{Interacting Bose gas in an inhomogeneous optical lattice}

\label{sec:inter}

In this section, we discuss the more general and practical case where the
atoms in an inhomogeneous optical lattice has collisional interactions with
each other. Many qualitative features discussed in the last section,
however, remain valid in the interacting case. The atomic interaction indeed
brings up several different properties in a quantitative level. In this
section, the emphasis of our discussion is on these differences.

By taking the interaction into account, the time-of-flight images are
expected to be modified in two major aspects. First, the repulsive
interaction between atoms will tend to broaden their spatial distribution in
a trap, and hence narrow the corresponding momentum distribution. Second,
during the time-of-flight expansion, the remnant atomic interaction
transforms the interaction energy into the kinetic energy, in particular in
the early stage of expansion. As a consequence, the momentum distribution
tends to be wider in the final images. The images also get a bit blurred due
to the scattering of different interference peaks in the momentum space
during the expansion. In this section, we will discuss both of these two
points.

For an interacting Bose gas in an inhomogeneous optical lattice, exact
solutions as in the non-interacting case no longer exist. Instead, we use
the local density approximation (LDA) to treat the inhomogeneity induced by
the global trap. When the interaction energy is much larger than the trap
energy, each local region of the global trap behaves like a homogeneous
system with interacting atoms in a pure optical lattice. The interaction in
this local homogeneous lattice is then analyzed under the
Hartree-Fock-Bogoliubov-Popov (HFBP) scheme~\cite{andersen-04, griffin-98}.
The validity of this approach is supported by the following considerations.
First, the LDA works well for a large number of atoms in a weak harmonic
trap, which is the case for the parameter ranges considered below. Second,
the HFBP method should be able to provide a reliable description of weakly
interacting Bose systems, except for the region close to a phase transition
(associated with the BEC or the Mott transition). For an atomic gas in a
global harmonic trap away from the Mott transition, such questionable region
corresponds only to a thin shell in space, and its contribution to the
global properties is far less significant.

Under the LDA, we consider an interacting Bose system with a spatially
dependent Hamiltonian %
\begin{equation}
H=\sum_{\mathbf{k}}(\epsilon _{\mathbf{k}}-\mu )a_{\mathbf{k}}^{\dagger } a_{%
\mathbf{k}}+\frac{U}{2}\sum_{\mathbf{k},\mathbf{k}^{\prime },\mathbf{q}} a_{%
\mathbf{k}+\mathbf{q}}^{\dagger }a_{-\mathbf{k}}^{\dagger } a_{\mathbf{k}%
^{\prime }+\mathbf{q}}a_{-\mathbf{k}^{\prime }},  \label{eqn:Hamiltonian-int}
\end{equation}
where $\epsilon_{\mathbf{k}}$ is the dispersion relation defined above, $\mu
\equiv \mu (\mathbf{r})$ is the local chemical potential, and $U\equiv U_{%
\mathrm{bg}}\int |w(\mathbf{r})|^{4}d^{3}\mathbf{r}$ is the on-site
interaction rate. For a typical experiment of $^{87}$Rb, $U_{\mathrm{bg}}$
is related to the $s$-wave background scattering length $a_{s}=5.45$ nm by $%
U_{\mathrm{bg}}=4\pi\hbar ^{2}a_{s}/m$, and $U$ takes an approximate form of
$U\approx3.05V_{0}^{0.85}(a_{s}/d)$, with an energy unit of the recoil
energy $E_{\mathrm{R}}$~\cite{gerbier-05}. With the standard HFBP approach,
we separate the bosonic operators into two parts: %
\begin{equation}
a_{\mathbf{k}}=\psi _{0}+\delta _{\mathbf{k}};\quad \quad a_{\mathbf{k}%
}^{\dagger }=\psi _{0}+\delta _{\mathbf{k}}^{\dagger },  \label{eqn:bosonic}
\end{equation}
where $\psi _{0}\equiv \langle a_{\mathbf{0}}^{\dagger }\rangle \equiv
\langle a_{\mathbf{0}}\rangle $ represents the condensate component, and $%
\delta _{\mathbf{k}}$ is the fluctuation around it. After performing the
substitution for $a_{\mathbf{k}}$ and $a_{\mathbf{k}}^{\dagger}$ into the
original Hamiltonian Eq. (\ref{eqn:Hamiltonian-int}), terms that are cubic
and quartic in $\delta _{\mathbf{k}}^{\dagger }$ and $\delta _{\mathbf{k}}$
will be present. These terms are reduced to quadratic forms under the HFBP
by employing the Wick's theorem. As a result, we obtain a quadratic
effective Hamiltonian $H_{\mathrm{eff}}$%
\begin{eqnarray}
H_{\mathrm{eff}} &\approx &\left( \epsilon _{0}-\mu +\frac{Un_{0}}{2}\right)
n_{0}  \notag  \label{eqn:Hamiltonian-int2} \\
&&+\sum_{\mathbf{k}}\left[ \epsilon _{\mathbf{k}}-\mu +2U(n_{\mathrm{tot}%
}-n_{0})\right] \delta _{\mathbf{k}}^{\dagger } \delta _{\mathbf{k}}  \notag
\\
&&+\frac{Un_{0}}{2}\sum_{\mathbf{k}}\left( \delta _{\mathbf{k}}^{\dagger}
\delta _{-\mathbf{k}}^{\dagger }+\delta _{\mathbf{k}}\delta _{-\mathbf{k}}
+4\delta _{\mathbf{k}}^{\dagger }\delta _{\mathbf{k}}\right).
\end{eqnarray}
Here, $\epsilon_{0}=-6t$ is the energy at the band bottom with $t$ is the
tunnelling rate defined above, $n_{0}=\psi _{0}^{2}$ is the per site density
of the condensate fraction, and $n_{\mathrm{tot}}$ is the total number of
particles per site. In order to derive the expression above, the saddle
point condition is employed to make the coefficients of terms linear in $%
\delta $ vanish, leading to the saddle point equation %
\begin{equation}
\mu (\mathbf{r})=\epsilon _{0}-Un_{0}+2Un_{\mathrm{tot}}.
\label{eqn:saddlept}
\end{equation}
This equation must be solved self-consistently with the number constraint $%
n_{\mathrm{tot}}=-\partial \Omega /\partial \mu $, where $\Omega
=-(1/\beta)\ln \mathrm{Tr}(e^{\beta H_{\mathrm{eff}}})$ is the
thermodynamical potential. This constraint leads to the number equation %
\begin{equation}
n_{\mathrm{tot}}=n_{0}+\sum_{\mathbf{k}\neq \mathbf{0}}\frac{1}{2} \left[
\frac{\epsilon _{\mathbf{k}}-\epsilon _{0}+Un_{0}}{E_{\mathbf{k}}} \coth
\left( \frac{\beta E_{\mathbf{k}}}{2}\right) -1\right] ,  \label{eqn:number}
\end{equation}
where $E_{\mathbf{k}}=\sqrt{(\epsilon _{\mathbf{k}}-\epsilon_{0}
+Un_{0})^{2}-U^{2}n_{0}^{2}}$ is the quasiparticle dispersion relation.

By fixing the number density per site at the trap center, we can solve Eqs. (%
\ref{eqn:saddlept}) and (\ref{eqn:number}) self-consistently to obtain the
chemical potential at the trap center $\mu_{0}$. This result, together with
the LDA relation of $\mu (\mathbf{r})=\mu_{0}-V(\mathbf{r})$, allows us to
calculate $\mu (\mathbf{r})$, and hence the condensate fraction $n_{0}(%
\mathbf{r})$ and quasi-momentum distribution of the non-condensate part $n_{%
\mathbf{k}\neq 0}(\mathbf{r})$ at arbitrary position in the trap. The
overall non-condensate quasi-momentum distribution, thus can be obtained by
integrating over the whole global trap. For the condensate component, it
should be emphasized that $n_{0}$ leads to a delta function at zero
momentum, which is an artificial result of LDA. In order to overcome this
artifact, one needs to consider explicitly the broadening of the condensate
momentum distribution due to the presence of the harmonic trap. This can be
done by the following procedure. First, we get the condensate fraction
distribution $n_{0}(\mathbf{r})$ over the trap. The condensate wave function
then can be well approximated by $\psi _{0}(\mathbf{r})=\sqrt{n_{0}(\mathbf{r%
})}$ under the Thomas-Fermi approximation~\cite{dalfovo-99}. The condensate
component of the quasi-momentum distribution is thus given by the Fourier
transform of the wave function $\psi _{0}(\mathbf{r})$. Second, by adding
this condensate contribution to that from the normal part, and multiplied by
the Wannier function square $|w(\mathbf{k})|^{2}$ as in Eq. (\ref%
{eqn:mom-dis-non}), we get the resulting momentum-space density distribution.

Next, as in the case of an ideal Bose gas, we discuss several
characteristics on the momentum-space density distribution. In Sec.~\ref%
{sec:vis-inter} and \ref{sec:distribution-inter}, we discuss the influence
of interaction on the atomic momentum distribution inside the trap,
especially the associated visibility and the peak width, respectively. Then,
we calculate in Sec.~\ref{sec:frac-inter} the condensate fraction as a
quantitative measure of the system temperature. Lastly, in Sec.~\ref%
{sec:expan-inter}, we analyze the interaction effect during the
time-of-flight expansion, and conclude that this effect does not change much
the characteristics discussed above.

\subsection{Visibility of the interference pattern}

\label{sec:vis-inter}

\begin{figure}[tbp]
\begin{center}
\includegraphics[width=8.0cm]{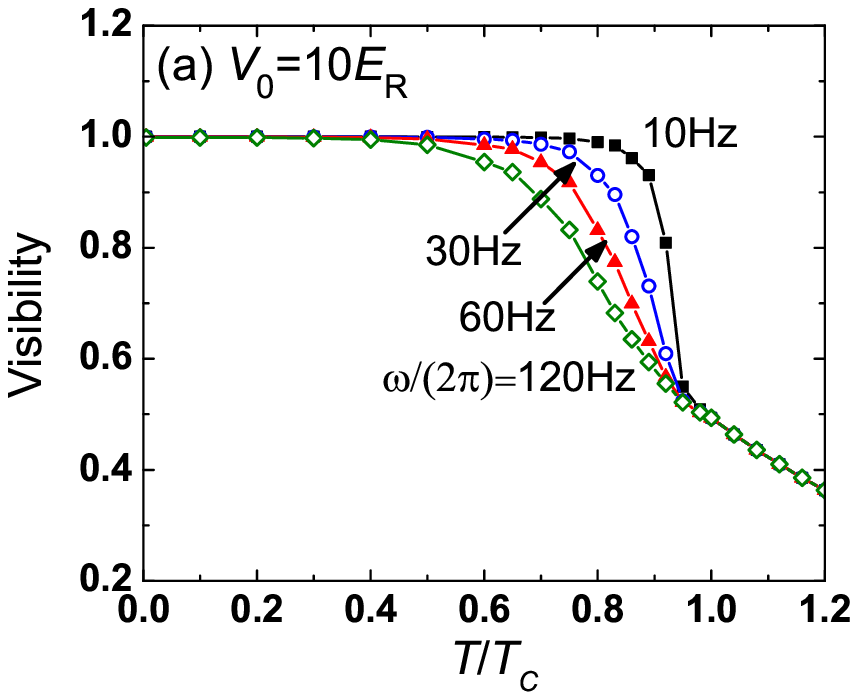} %
\includegraphics[width=8.0cm]{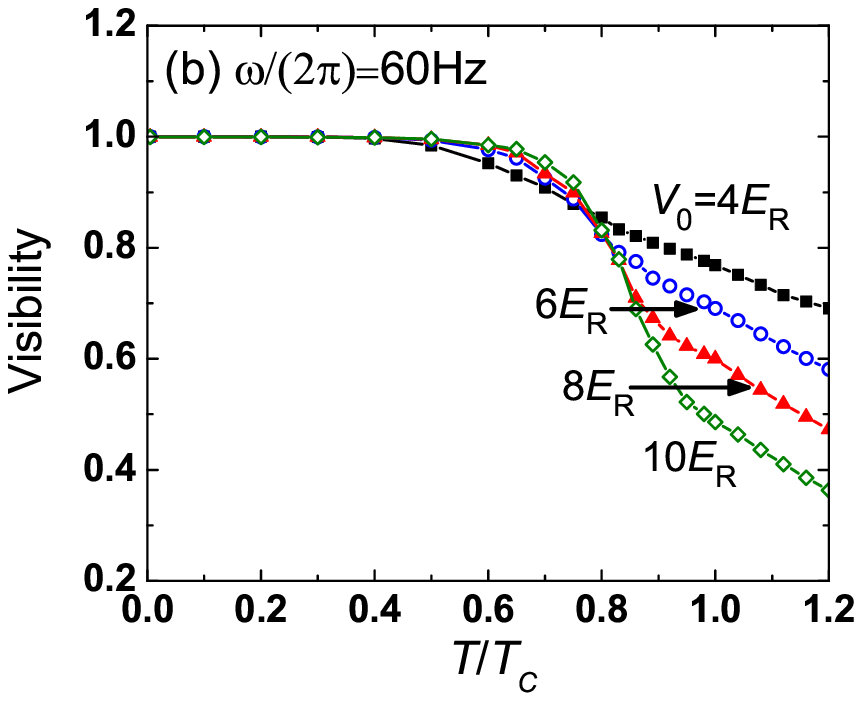}
\end{center}
\caption{(color online) The visibility as a function of temperature around the BEC
transition temperature $T_{c}$ for (a) $V_{0}=10E_{\rm R}$ and (b) $\protect%
\omega =2\protect\pi \times 60$ Hz. The number density per site at the trap
center is set as $n_{\mathrm{tot}}=1$, and the total number of particles in
the trap is $N \sim 10^{4}$. }
\label{fig:visi-inter}
\end{figure}

We show in Fig.~\ref{fig:visi-inter} the visibility of the interference
pattern as a function of temperature across the BEC transition. Since the
presence of interaction sets another energy scale, the system is not
determined by the single parameter of $\hbar ^{2}\omega ^{2}/(tE_{\mathrm{R}%
})$, as in the ideal gas case. Instead, we consider various combinations of
lattice barriers and trapping frequencies. %

>From Fig.~\ref{fig:visi-inter}(a), we find that independent of the strength
of the global trap, the thermal visibility with $T \gtrsim T_{c}$ remains
small (with $v$ about or below $0.4$). This result is significantly
different from the case of an ideal Bose gas, where the thermal visibility
can be close to the unity for a very weak global trap. By crossing the BEC
transition, the visibility clearly increases so that for an interacting Bose
gas, a high visibility signifies that the atomic cloud is in the condensate
region. However, for a strong global trap (with $\omega \sim 2\pi \times 120$
Hz for instance), the variation of the visibility is not sharp at the
transition point, but continues into a pretty wide region below $T_{c}$.
Thus, in this case, it becomes less accurate to use the jump of the
visibility to determine the BEC transition temperature.

Another feature we can read from Fig.~\ref{fig:visi-inter}(a) is the
convergent behavior of visibility for a thermal gas with the same lattice
potential but various values of trapping frequency. This behavior can be
understood from the overall quasi-momentum distribution %
\begin{equation}
n_{\mathbf{k}}=\int d^{3}\mathbf{r}n_{\mathbf{k}}(\mathbf{r}) = 4\pi
\int_{0}^{\infty }drn_{\mathbf{k}}(r)r^{2},  \label{eqn:mom-dis-inter}
\end{equation}
where the integration is over the whole trap. Under the LDA, the spatial
dependence of number density is only through the chemical potential $n_{%
\mathbf{k}}(\mathbf{r})={\tilde{n}}_{\mathbf{k}}(\mu (\mathbf{r}))$, hence
the integration can be rewritten as %
\begin{equation}
n_{\mathbf{k}}=4\pi \left( \frac{m}{2\omega ^{2}}\right) ^{3/2}
\int_{\mu_{0}}^{0}\varepsilon {\tilde{n}}_{\mathbf{k}}(\varepsilon ) d\sqrt{%
\varepsilon }.  \label{eqn:mom-dis-inter2}
\end{equation}
For a thermal gas with a certain number density and chemical potential at
the trap center, the function ${\tilde{n}}_{\mathbf{k}}(\varepsilon )$, and
hence the integration over $\varepsilon $ in the equation above is fixed.
Thus, all the trap can do is to re-scale the quasi-momentum distribution by
a factor of $\omega ^{-3}$. For a given optical lattice characterized by a
Wannier function, the momentum-space density profile of a thermal gas for
various values of $\omega $ takes the identical shape, hence all signatures
we can read from it must remain the same. Notice that since the LDA approach
is reliable for all parameter ranges discussed here (with $\omega \lesssim
2\pi \times 120$ Hz), we conclude that this result is the effect of a strong
interaction compared to the trapping potential, which can not be smoothly
connected to the non-interacting results.

In Fig.~\ref{fig:visi-inter}(b), we show the visibility for different
lattice depths with a fixed strength of the global trap. With lower lattice
depths, the thermal visibility increases. For $V_{0}=4E_{\mathrm{R}}$, for
instance, the visibility varies almost linearly with temperature near the
condensation transition (for $T$ from $0.5T_{c}$ to $1.2T_{c}$), and a high
thermal visibility remains (with $v \sim 0.7-0.8$) even when $T$ crosses $%
T_{c}$. This flat behavior hence makes it difficult to use visibility to
signify the condensate region and to identify the transition point for
interacting atoms in a shallow lattice.

\subsection{Momentum-space density profile and the peak width}

\label{sec:distribution-inter}

\begin{figure}[tbp]
\begin{center}
\includegraphics[width=8.0cm]{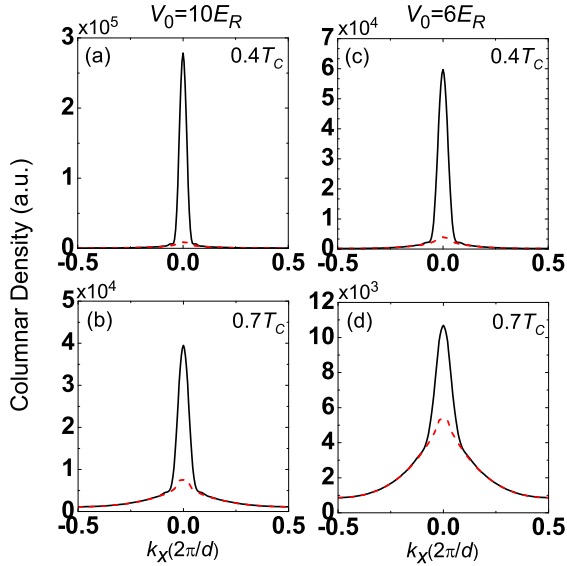}
\end{center}
\caption{(color online) Momentum space columnar density along the $x$-axis in the first
Brillouin zone, where clear bimodal structures appear for temperatures below $%
T_{c}$. Here, the solid curves are the total momentum density profile, and
the dashed curves are the momentum density profile of the normal component.
The trapping frequency used in these plots is $\protect\omega =2 \protect\pi %
\times 60$ Hz, and the number density per site is unity at the trap center. }
\label{fig:distri-inter}
\end{figure}

After discussing the visibility of interference peaks, we next focus on the
momentum distribution in the first Brillouin zone. In Fig.~\ref%
{fig:distri-inter}, we show the column-integrated momentum density profile
along one of the crystallographic axis (say, the $x$-axis). Similar to the
case of an ideal Bose gas, a clear bimodal structure appears for
temperatures below $T_{c}$ even when the condensate fraction is still small
[see Fig.~\ref{fig:distri-inter}(b) with $n_{0}\sim 9\%$]. Therefore, the
momentum distribution and its bimodal structure sets an unambiguous
criterion for the BEC transition, especially when supplemented with the
interference pattern from the lattice structure. %

\begin{figure}[tbp]
\begin{center}
\includegraphics[width=8.0cm]{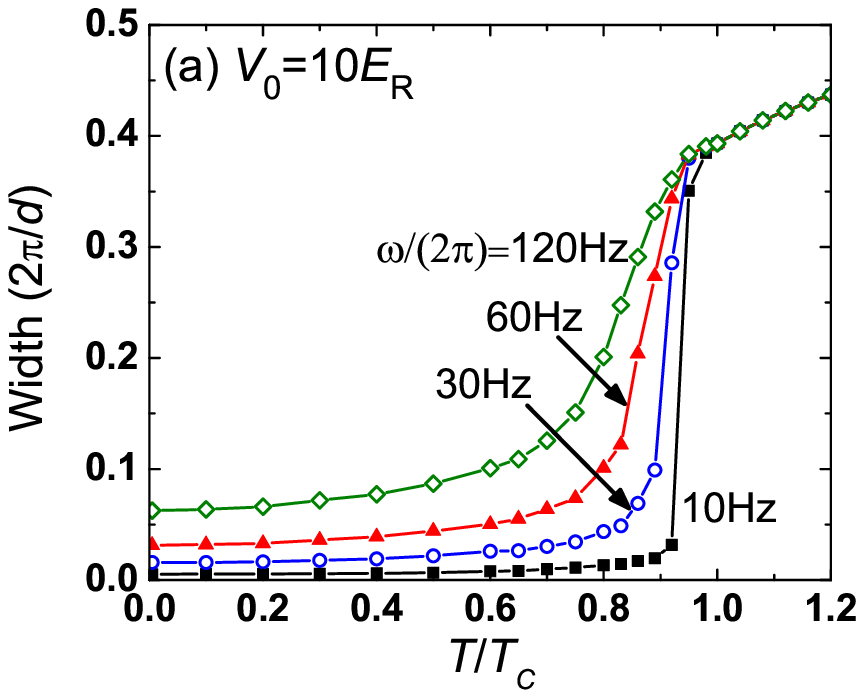}
\includegraphics[width=8.0cm]{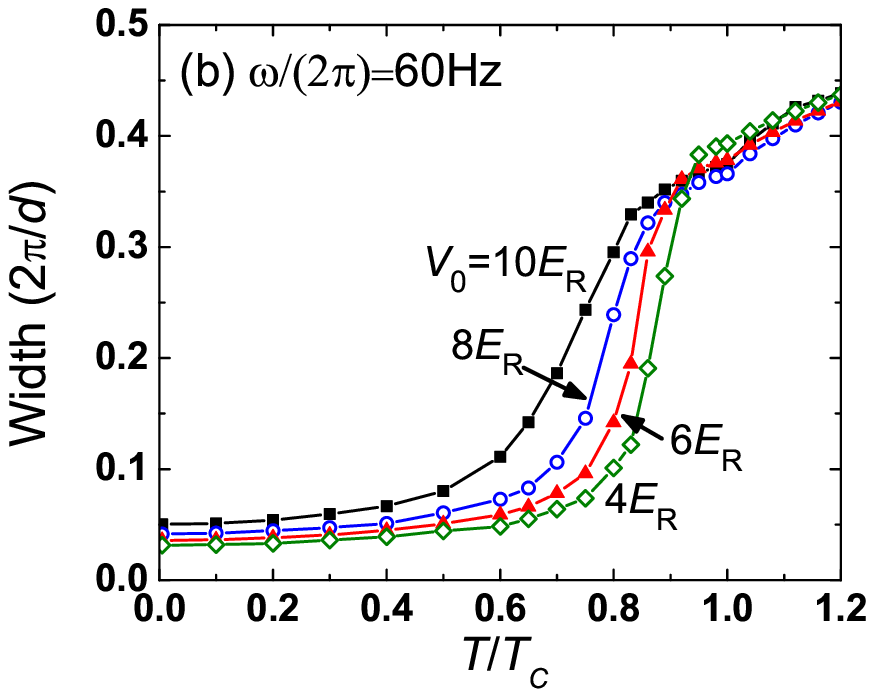}
\end{center}
\caption{(color online) The peak width within the first Brillouin zone, taking across the
transition temperature for (a) $V_{0}=10 E_{\mathrm{R}}$ and (b) $\protect%
\omega =2\protect\pi \times 60$ Hz. The number density per site is unity at
the trap center. }
\label{fig:width-inter}
\end{figure}

To characterize the sharpness of the central peak, we still use the peak
width defined above as a single-value parameter. As shown in Fig.~\ref%
{fig:width-inter}, the peak width significantly reduces across the BEC
transition, so a sharp peak with a small width sets a clear indicator that
the system is in the condensate region. This conclusion is qualitatively
consistent with the case of an ideal Bose gas as discussed in Sec.~\ref%
{sec:distribution-non}. For an ideal Bose gas, the peak width always has a
sharp and large jump at the BEC transition point. For an interacting Bose
gas, this jump becomes less sharp under certain circumstances. From Fig.~\ref%
{fig:width-inter}, we notice that the jump of the peak width remains sharp
across $T_{c}$ for weak global traps or for lower lattice depths. However,
in a stronger global trap with a higher lattice depth, the decrease of the
peak width takes place over a wider range of temperatures, which makes it
less accurate to locate the transition point using the peak width as an
indicator. If one compares Figs. \ref{fig:visi-inter} and \ref{fig:width-inter},
it is interesting to note for that for a weak optical lattice,
even when the visibility becomes too flat to show a phase transition,
the variation of the peak width remains significant and sharp to serve as
an indicator of the condensation transition. %

\subsection{The condensate fraction}

\label{sec:frac-inter}

\begin{figure}[tbp]
\begin{center}
\includegraphics[width=8.0cm]{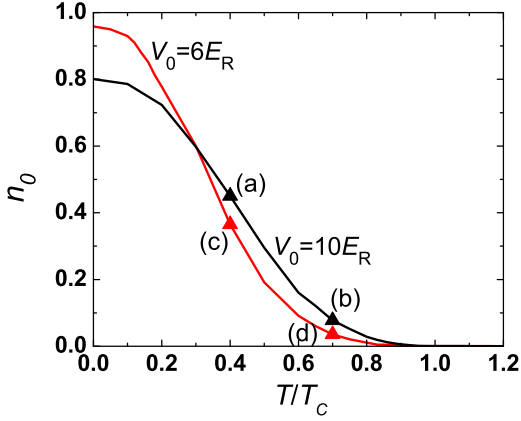}
\end{center}
\caption{(color online) The condensate fraction as a function of temperature below the BEC
transition, for an interacting Bose gas in an optical lattice with $\protect%
\omega =2\protect\pi \times 60$ Hz and different lattice barriers $V_{0}$.
The number density per site is unity at the trap center. By comparing with
the momentum distribution as shown in Fig.~\protect\ref{fig:distri-inter},
temperatures of such a system can be determined correspondingly (triangles).
}
\label{fig:n0-inter}
\end{figure}

\begin{figure}[tbp]
 \begin{center}
  \includegraphics[width=8.0cm]{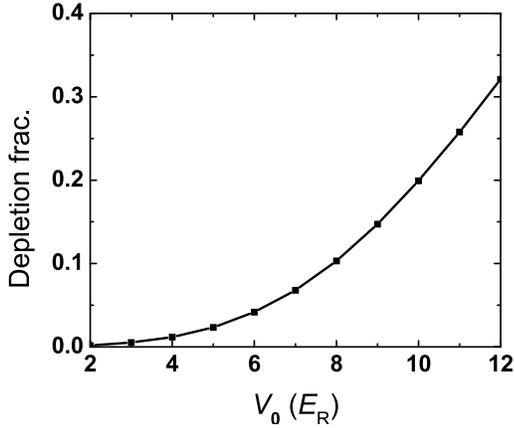}
 \end{center}
\caption{ The in-trap quantum depletion fraction as as a function
of the lattice barrier $V_0$. The trapping frequency is $\omega= 2\pi\times 60$ Hz.
The number density per site at the trap center is unity.}
\label{fig:qd}
\end{figure}

As in the case of an ideal Bose gas, when one moves into the condensate
region, the visibility and the peak width become insensitive to the
variation of temperature. Instead, we need to use the condensate fraction as
an indicator of the temperature in the condensate region. The condensate
fraction can be measured similarly through the bimodal fitting to the atomic
momentum distribution. In Fig.~\ref{fig:n0-inter}, we show the total
condensate fraction as a function of the temperature under a couple of
different lattice barriers. First, the condensate fraction change
monotonically and sensitively with temperature, so it provides a potentially
good thermometer. Second, it is interesting to note that the condensate
fraction does not approach the unity even when the temperature tends to
zero. This result is significantly different from the case of an ideal Bose
gas, where at zero temperature it is always a pure condensate. This
discrepancy is due to quantum depletion of the condensate at zero
temperature, which is always present for an interacting gas. In the zero
temperature limit, the HFBP approximation used here reduces to the
Bogoliubov approach, so it naturally takes into account the contribution
from quantum depletion.
In Fig.~\ref{fig:qd}, we show the in-trap zero temperature quantum
depletion fraction for various lattice barrier depths. Notice that
by applying a higher lattice barrier $V_0$, the quantum depletion
fraction gets more significant. This observation justifies
the zero temperature results of condensate fraction as shown in
Fig.~\ref{fig:n0-inter}, and is consistent with the trend one should expect.
In fact, in the case with no optical lattice, the condensate fraction
should approach unity at zero temperature for a weakly interacting dilute
gas (with a small gas parameter). In the opposite limit, when the lattice
barrier tends to the critical value for the Mott transition, the condensate
fraction should deplete to zero. When the quantum depletion is dramatic,
the HFBP is no longer a good approximation. However, for parameter ranges
discussed above, even with $V_{0}=12E_{\mathrm{R}}$, the condensate fraction
($\sim 70\%$) still dominates at zero temperature, assuring the validity of
the HFBP approximation used throughout this section.
\begin{figure}[tbp]
 \begin{center}
  \includegraphics[width=8.0cm]{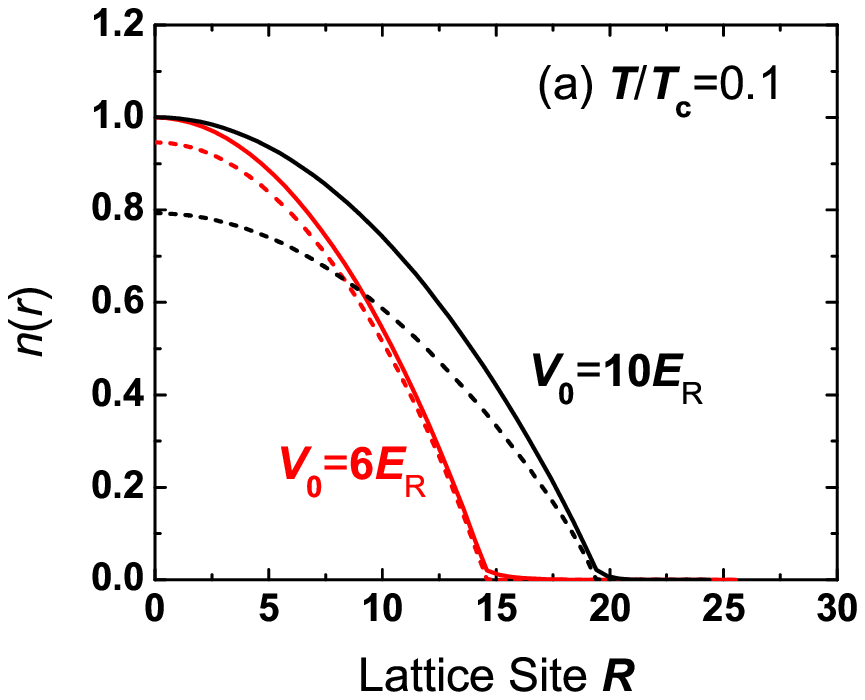}
  \includegraphics[width=8.0cm]{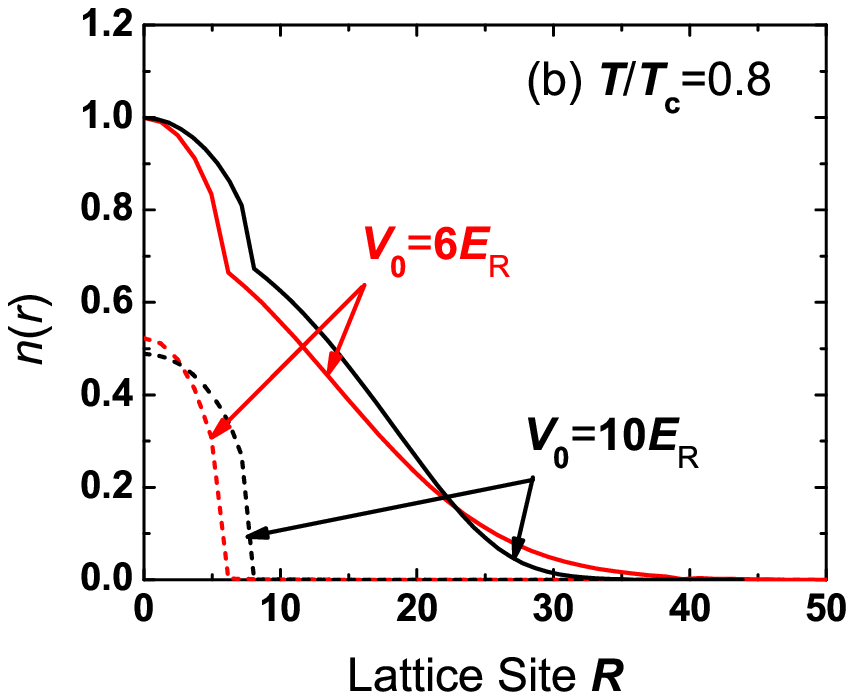}
 \end{center}
\caption{(color online) The number and condensate density distribution
along the radial direction away from the trap center.
Two different temperatures are considered as in (a) $T=0.1T_c$, and
(b) $T=0.8T_c$. The solid curves correspond to the overall density profiles
and the dashed curves to the condensate portions. The trapping frequency
used here is $\omega=2\pi\times 60$ Hz. The total number of particles
$N_\mathrm{tot}$ and the number of condensate particles $N_0$ are respectively
(a) $V_0=6E_\mathrm{R}$(red): $N_\mathrm{tot}=0.54\times 10^4$, $N_0=0.50\times 10^4$;
$V_0=10E_\mathrm{R}$(black): $N_\mathrm{tot}=1.3\times 10^4$, $N_0=1.0\times 10^4$;
(b) $V_0=6E_\mathrm{R}$(red): $N_\mathrm{tot}=2.5\times 10^4$, $N_0=2.8\times 10^2$;
$V_0=10E_\mathrm{R}$ (black): $N_\mathrm{tot}=2.3\times 10^4$, $N_0=6.7\times 10^2$.}
\label{fig:intrap}
\end{figure}

Another interesting feature of the condensate fraction plot Fig.~\ref{fig:n0-inter}
is the crossing behavior of the two curves for different $V_0$ at finite temperature.
For a homogeneous gas with a certain number density, one would expect a more severe
depletion (both quantum and thermal) from the condensate for the higher $V_0$ case,
since both temperature and interaction would play relatively more significant role
when the transfer integral $t$ decreases. However, this simple trend becomes more
complicated in a trapped case. In fact, due to the presence of a global trap, thermal
wings must emerge at the edge at any finite temperature. As one calculate
the in-trap condensate fraction, an integration over the whole trap should take the
thermal wings into account, and the result will inevitably depend on how much
the thermal part contribute to the total number. As an example, we show the in-trap
number and condensate density profiles in Fig.~\ref{fig:intrap} with fixed total number
density at the trap center. At a lower temperature $T=0.1T_c$, the thermal contribution
is negligible so it is apparent that a higher depletion is observed in higher barriers,
as shown in Fig.~\ref{fig:intrap}(a). At a higher temperature $T=0.8 T_c$,
the thermal part becomes significant and one observes a more extended thermal tail
in the lower barrier case with $V_0=6E_{\rm R}$, as shown in Fig.~\ref{fig:intrap}(b).
This greater thermal contribution to the total number makes the condensate part
relatively smaller and leads to a smaller condensate fraction.
Therefore, the crossing behavior of $n_0$ as shown in Fig.~\ref{fig:n0-inter}
is a consequence of the inhomogeneous distribution of different phases with various
number density and chemical potential, which are results from the inhomogeneity of
the trap.

\subsection{Interaction effects during the time-of-flight expansion}

\label{sec:expan-inter}

Up to now, we have calculated the atomic momentum distribution inside the
trap and have neglected modification of this distribution caused by the
atomic interaction during the time-of-flight expansion. During the expansion
of the atomic cloud, the initial interaction energy is transferred to the
kinetic energy, which tends to broaden the momentum distribution \cite{note}%
. This influence is most evident for the condensate part, due to its high
number density and narrow momentum distribution at the beginning. The
influence on the momentum distribution of the thermal and the non-condensate
part is negligible, as confirmed by experiments~\cite{gerbier-07}, since
this part has a broad initial momentum distribution already, and the weak
atomic collisions are unlikely to cause any significant modification. In the
following, we only discuss influence of the atomic collisions on the
condensate part of the momentum distribution.

After turnoff of the optical lattice and the confining potential, the atomic
cloud undergoes a free expansion in space, and the expansion of the
condensate part can be well described by a time dependent Gross-Pitaevskii
(TDGP) equation %
\begin{equation}
i\hbar \partial _{t}\Psi (\mathbf{r},t)= \left(-\frac{\hbar ^{2}\nabla_{%
\mathbf{r}}^{2}}{2m} + U|\Psi (\mathbf{r},t)|^{2}\right) \Psi (\mathbf{r},t),
\label{eqn:TDGP}
\end{equation}
where the initial condition $\Psi (\mathbf{r},t=0)$ ($t$ is the expansion
time) is given by the equilibrium condensate wave function inside the
optical lattice and global trap, which has been calculated with the method
detailed in the above. Note that only for the condensate part we use the
TDGP to evolve its momentum distribution.

It is easier to understand the consequence of this evolution by looking at
the TDGP equation in the momentum space. The Fourier transform of Eq. (\ref%
{eqn:TDGP}) gives %
\begin{eqnarray}
i\hbar \partial _{t}\Psi _{\mathbf{k}}(t) &=&\frac{\hbar ^{2}k^{2}}{2m} \Psi
_{\mathbf{k}}(t)  \notag  \label{eqn:TDGP2} \\
&&\hspace{-1.5cm}+U\sum_{\mathbf{k}^{\prime },\mathbf{q}} \Psi _{-\mathbf{k}+%
\mathbf{q}}^{\dagger }(t)\Psi _{-\mathbf{k}^{\prime } +\mathbf{q}}(t)\Psi _{%
\mathbf{k}^{\prime }}(t).
\end{eqnarray}
Clearly, without atomic collisions (the $U$ term), the momentum distribution
$|\Psi _{\mathbf{k}}(t)|^{2}$ remains unchanged. The atomic collisions
transfer a pair of atoms from momenta ($\mathbf{k}^{\prime },-\mathbf{k}%
^{\prime }+\mathbf{q}$) to ($\mathbf{k},-\mathbf{k}+\mathbf{q}$), which
modulates the overall momentum distribution. %
\begin{figure}[tbp]
\begin{center}
\includegraphics[width=6.0cm]{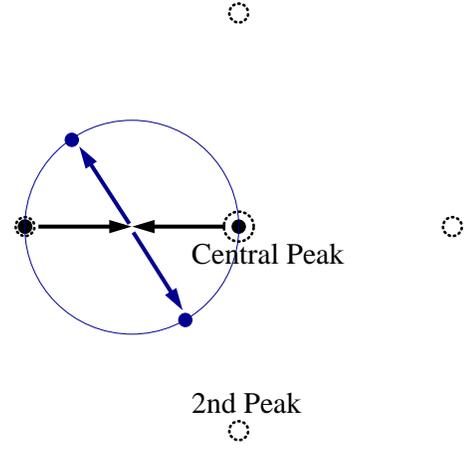}
\end{center}
\caption{ Illustration of the scattering between atoms from different peaks
(the central and the secondary). This process leads to blurring of the
interference peaks with a characteristic blurring pattern. }
\label{fig:scatter}
\end{figure}

To understand the consequence of the collision induced momentum transfer, we
note that the initial condensate momentum distribution has a number of
peaks, one in each Brillouin zone. As the central peak in the first
Brillouin zone (with $\mathbf{k}$ close to zero) is the highest one, the
scattering of pairs of atoms around the first peak satisfying the momentum
conservation has the largest contribution to the collision effect. Around
the central peak, the wave function $\Psi _{\mathbf{k}}(t)$ has approximate
spherical symmetry, and the evolution of the momentum distribution around
the central peak by the TDGP equation has been calculated and shown in Ref.~%
\cite{yi-07}. The condensate peak gets somewhat lower and broader, however,
its width is typically still significantly less than the width of the
thermal cloud, and a bimodal structure of the momentum distribution remains
clearly visible. Thus, the collision during the time-of-flight expansion has
some quantitative influence on the peak width we calculate before, but the
effect is not large and it should not change all the qualitative discussions
in the last sections. In particular, as the bimodal structure remains
clearly observable, we do not expect that the condensate fraction measured
through the bimodal fitting to the momentum distribution has any significant
change by the collision effect during the expansion.

In the next order, the atom in the central peak can collide with another
atom in the secondary peak, and scatter to some other directions in the
momentum space. The momentum difference between the central peak and the
secondary peak is given by $\mathbf{G}$ ($\mathbf{G}$ is related to the
lattice constant $d$ through $|\mathbf{G}|=2\pi /d$), so the kinetic energy
difference between them corresponds to a large energy scale $\hbar ^{2}|%
\mathbf{G}|^{2}/(2m)$, which is typically larger than the interaction
strength during the expansion [the latter can be estimated by $%
U n\left(t\right) $, where $n\left( t\right) $ is the instantaneous atomic
density at the collision]. Therefore, to be effective, the collisions need
to satisfy the momentum conservation as well as an approximate energy
conservation in the momentum space. As a consequence, for atoms with
incoming momenta around $0$ and $G$ (corresponding to the central and the
secondary peaks, respectively), the outgoing atoms are centered around a
spherical surface in the momentum space as shown in Fig.~\ref{fig:scatter}
[with momenta $\left( \mathbf{k}^{\prime },-\mathbf{k}^{\prime }+\mathbf{G}%
\right) $, where $\mathbf{k}^{\prime 2}+\left( -\mathbf{k}^{\prime }+\mathbf{%
G}\right) ^{2}\approx |\mathbf{G}|^{2}$]. The sphere has origin at
$\mathbf{k}=\mathbf{G}/2$ and a radius of $|\mathbf{G}|/2$. This collision effect causes some
blurring of the original peaks. Since the scattered atoms are dominantly
around a sphere in the momentum space, this blurring scattering gives some
characteristic momentum distribution pattern. Experimentally, by looking at
such a pattern, one may measure and constraint the magnitude of collision
effects during the time-of-flight expansion.

At high orders, there could be also scattering between different secondary
peaks as well as scattering between the central and even higher order peaks.
Although these scattering can change some of quantities we calculate before,
we expect all these modifications are pretty small.

\section{Conclusion}

\label{sec:conclusions}

In summary, we have discussed in this manuscript ideal as well as
interacting Bose gases in an inhomogeneous optical lattice within a global
harmonic trap. By explicitly calculating the momentum distribution, we have
studied several possible signatures of the BEC transition in an lattice
based on the common detection technique of time-of-flight imaging. For
parameters of relevance to the current experiments, a large visibility, a
substantial decline of the peak width, and the appearance of a bimodal
structure for the central peak, can all be used as signals of the
condensation transition as one decreases the temperature. For some other
parameters, the thermal visibility could be significant, and in such a case
the latter two criteria will work better. In particular, the appearance of a
bimodal structure for the momentum distribution is a robust signal
associated with the condensation transition in both free space and lattice
(in the lattice case, the interference peaks give further information about
the underlying lattice structure).

After the condensation transition, both the visibility and the peak width
become insensitive to the variation of temperature, so they can not serve as
a practical thermometer. Instead, one may measure the condensate fraction by
a bimodal fitting to the atomic momentum distribution. The condensate
fraction changes steadily with temperature, and may work as a good
experimental indicator of the system temperature by comparing with the
results from theoretical calculations.

\begin{acknowledgements}

We thank Immanuel Bloch for helpful discussions. This work is
supported under ARO Award W911NF0710576 with funds from the DARPA
OLE Program and the MURI program.

\end{acknowledgements}



\begin{references}
\bibitem{note1}
The cases of different lattice structures and anisotropic trapping
potentials can be analyzed analogously, provided that the
anisotropy of the global trap is not strong enough to cause a
breakdown of LDA.
\end{references}
\end{document}